\documentclass[twocolumn, english, aps, pra, showpacs, superscriptaddress]{revtex4-2}

\usepackage[T1]{fontenc}
\usepackage[latin9]{inputenc}
\usepackage{amsmath}
\usepackage{amssymb}
\usepackage{color}
\usepackage{times}
\usepackage{units}
\usepackage{amsmath}
\usepackage{amssymb}
\usepackage{graphicx}
\usepackage{wasysym}
\usepackage{soul} 
\usepackage{cancel}
\usepackage{babel}
\usepackage{hyperref}
\usepackage{dsfont}
\usepackage{lipsum}
\usepackage{braket}
\usepackage{orcidlink}

\def\equationautorefname~#1\null{Equation (#1)\null}



\begin{document}
\title{Magnetic Field Independent SERF Magnetometer}
\author{Mark Dikopoltsev}
\affiliation{Institute of Applied Physics, The Faculty of Science, The Center for Nanoscience and Nanotechnology, The Hebrew University of Jerusalem, Jerusalem 91904, Israel}
\affiliation{Rafael Ltd, 31021, Haifa, Israel.}

\author{Uriel Levy}
\affiliation{Institute of Applied Physics, The Faculty of Science, The Center for Nanoscience and Nanotechnology, The Hebrew University of Jerusalem, Jerusalem 91904, Israel}

\author{Or Katz}
\author{Or Katz\orcidlink{0000-0001-7634-1993}}
\email{or.katz@cornell.edu}
\affiliation{School of Applied and Engineering Physics, Cornell University, Ithaca, NY 14853.}

\begin{abstract}
SERF magnetometers based on dense ensembles of alkali-metal spins are precision quantum sensors that hold the record of measured and projected sensitivity to magnetic fields, in the $\mu\textrm{G}-\textrm{mG}$ range. At geomagnetic fields however, these sensors quickly lose their magnetic sensitivity due to spin decoherence by random spin-exchange collisions. Here we discover that atoms with nuclear spin $I=1/2$ can operate in the Spin-Exchange Relaxation Free (SERF) regime even at high magnetic field. We counter-intuitively show that frequent collisions between a dense and optically-inaccessible $(I=1/2)$ gas with another optically-accessible spin gas ($I>1/2$) improve the fundamental magnetic sensitivity of the latter. We analyze the performance of a dual-specie potassium and atomic hydrogen magnetometer, and project a fundamental sensitivity of about $10\,\mathrm{aT}\sqrt{\mathrm{cm}^3/\mathrm{Hz}}$ at geomagnetic fields for feasible experimental conditions. 

\end{abstract}

\maketitle
\section{Introduction}
Precision sensing of magnetic fields is a cardinal technique in various scientific disciplines \cite{jiang2021search,gentile2017optically,jensen2016non,hernandez2007ultrasensitive,oelsner2022integrated,bloch2022new,neukirch2010development,robinson2019developing,xu2006magnetic,ledbetter2008zero,bloch2023constraints,afach2021search,bloch2024rotating}. Warm ensembles of alkali-metal atoms feature an unprecedented sensitivity for measuring slowly varying and constant magnetic fields \cite{budker2007optical,kominis2003subfemtotesla,dang2010ultrahigh,ledbetter2008spin,gartman2018linear}. The high sensitivity of these sensors relies on the efficient coupling of the electron spin of each alkali-metal atom to the field and the use of a macroscopic ensemble that responds in a collective manner.

At geomagnetic field, the fundamental sensitivity of magnetometers based on alkali-metal atoms reaches a lower bound around $0.6\,\mathrm{fT}\sqrt{ \mathrm{cm}^3/\mathrm{Hz}}$. The dependence of this limit on the atomic number density is very weak \cite{smullin2009low}, because an increase in the number of spins which probe the field leads to increased decoherence by collisions \cite{happer1973spin}. Yet, At some conditions it is possible to suppress the decoherence caused by spin-exchange collisions, which are the dominant collision process in dense ensembles of alkali atoms in their electronic ground state \cite{happer1977effect,kominis2003subfemtotesla,korver2013suppression}. Such operation corresponds to the Spin-Exchange-Relaxation Free (SERF) regime, which features an improved fundamental sensitivity by about two orders-of-magnitudes, down to $10~\,\mathrm{aT}\sqrt{\mathrm{cm}^3/\mathrm{Hz}}$, where the atoms response is limited by other relaxation processes \cite{dang2010ultrahigh,jau2010optically}. To date, operation in this regime at low magnetic fields holds the record of best realized magnetic sensitivity \cite{allred2002high}.

Operation in the SERF regime however poses stringent conditions on the degree of spin-polarization of the atoms \cite{jau2004intense, smullin2009low, sheng2013subfemtotesla}, on the selectivity of the optical transitions \cite{Berrebi2022}, or on the magnitude of the ambient magnetic field \cite{kominis2003subfemtotesla, dang2010ultrahigh, ledbetter2008spin, balabas2010polarized,chalupczak2014spin, kong2020measurement, jimenez2014optically, katz2013nonlinear, chalupczak2014spin, xiao2021atomic, mouloudakis2022effects,Dikopoltsev2022}. Near unity polarized spins are less susceptible to spin-exchange relaxation but typically require high optical power at pulsed operation. Operation at low magnetic fields on the other hand, requires high atomic densities, and is practically limited to a few milligauss. The low-field requirement often  renders the SERF regime impractical for most applications which operate in a magnetically unshielded environment, in the presence of earth's magnetic field (a few hundreds of milligauss). 

Here we propose and analyze the operation of a hydrogen-based magnetometer which operates in the SERF regime and is unrestricted to low magnetic fields. We first discover that hydrogen atoms are inherently free of spin-exchange relaxation at any magnetic field or degree of spin-polarization, owing to their simple spin structure. We then analyze the operation of a hybrid potassium-hydrogen magnetometer, which enables control and measurement of the optically inaccessible spin-state of the hydrogen ensemble through its efficient collisional coupling.
We demonstrate that the potassium atoms inherit the coherent spin properties of the hydrogen, and can therefore attain an improved fundamental sensitivity down to $10\,\mathrm{aT}\sqrt{\mathrm{cm}^3/\mathrm{Hz}}$ at high magnetic fields. We finally outline an experimentally feasible configuration for realizing such a sensor.

\begin{figure}[t]
\begin{centering}
\includegraphics[width=8.4cm]{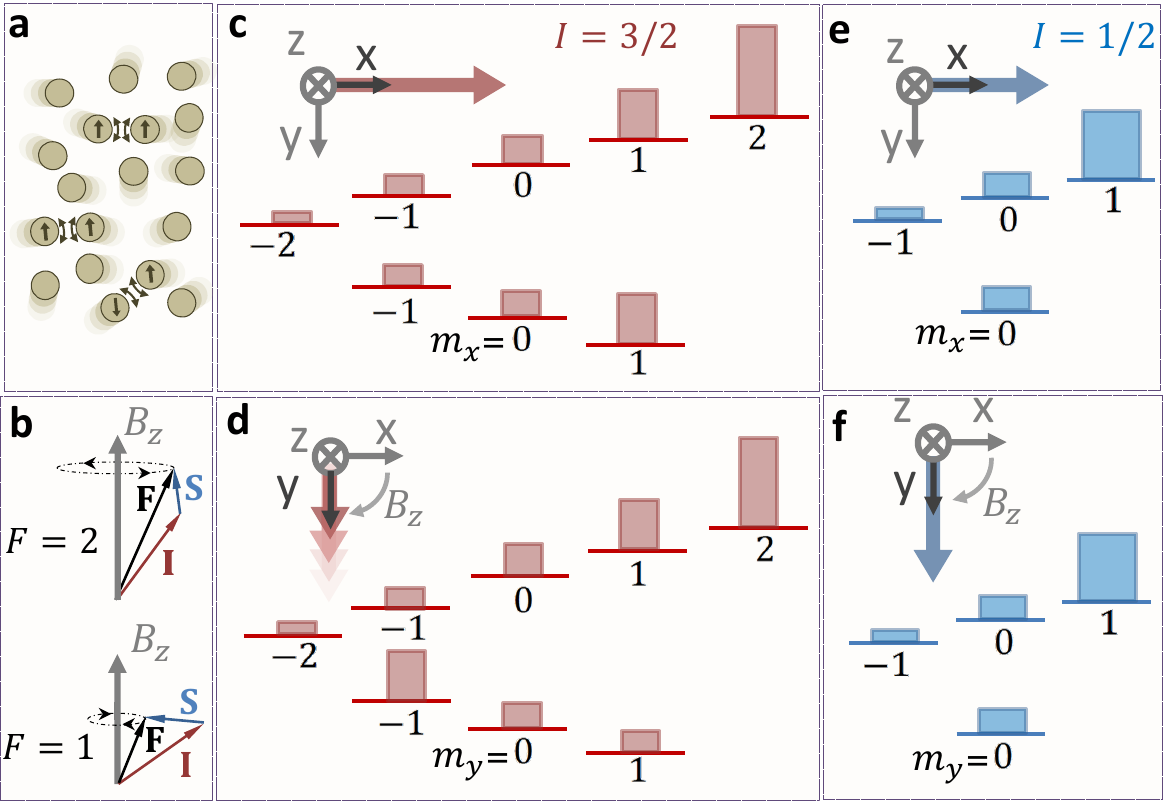}
\par\end{centering}
\centering{}\caption{\textbf{Spin-exchange relaxation} \textbf{a}, Frequent binary collisions of warm atomic gas lead to efficient exchange of spin. \textbf{b}, Alkali-metal atoms with a single valence electron are comprised of two hyperfine manifolds in the electronic ground-state. The total spin $\textbf{F}$ precesses clockwise (counter-clockwise) in the upper (lower) manifold in between sudden collisions. \textbf{c-d}, Spin-exchange relaxation mechanism for atoms with nuclear spin $I=3/2$. \textbf{c,} For spin ensembles oriented along $x$, the populations of the spin levels follow a spin temperature distribution (STD) $\rho\propto e^{\beta m_x}$, the equilibrium state by collisions. \textbf{d,} Spin-precession by the magnetic field along $z$ deters the STD, leading to decoherence by collisions; e.g.~after a $\pi/2$ pulse the spin distribution is $\rho\propto e^{(-1)^{\chi}\beta m_y}$ instead, where the population in the lower hyperfine manifold is inverted, see text. \textbf{e-f}, Precession of spin $1/2$ atoms maintains the spin-temperature distribution at all times, thus suppressing the relaxation associated with spin-exchange collisions. \textbf{c,e}, Distributions at $t=0$ with quantization axis along $\hat{x}$. \textbf{d,f}, Distributions at $t=\pi/(2\gamma B_z)$ with quantization axis along $\hat{y}$. \label{fig:energy_level}}
\end{figure}

\section{Dynamics of Ensembles in a Spin-Temperature Distribution}
The main limitation on the sensitivity of warm atomic magnetometers operating at high magnetic fields originates from relaxation induced by frequent spin-exchange collisions. In the mean-field (single-spin) picture \cite{anderson1959n,happer1972optical,happer1977effect, savukov2005effects,horowicz2021critical}, collisions predominantly alter the internal spin distribution of each atom, driving it toward a spin-temperature distribution (STD), which represents the steady-state of the spin-exchange interaction \cite{happer1977effect}. The ground-state energy levels of alkali-metal atoms, whose nuclear spin $I$ is nonzero, are split into two different hyperfine manifolds with quantum numbers $F,m$ (with $F=I\pm S$ and $|m|\leq F$) and have a density matrix $\rho(F,m)\propto e^{\beta m}$, following STD, as shown in Fig.~\ref{fig:energy_level}. The inverse temperature coefficient $\beta(P)$ depends on the degree of spin-polarization of the vapor $0\leq P\leq1$, which is a conserved quantity in the absence of other relaxation processes  \cite{appelt1998theory}. For this distribution and at ambient conditions, the two hyperfine manifolds are populated independently of the hyperfine energy splitting.

A magnetic field can perturb the spin-distribution of the atoms, and compete with spin-exchange collisions which act to maintain it at STD. In the presence of a magnetic field, the spins experience Larmor precession, whose rate varies between the two hyperfine manifolds. For atoms with $I>1/2$, the electron spin in the upper hyperfine manifold is aligned with the field, but is anti-aligned at the lower manifold. Consequently, the total spin in the two hyperfine manifolds precess at the same rate $\gamma=\gamma_e/(2I+1)$ but at opposite directions where $\gamma_e=28 \,\mathrm {MHz/mT}$ is the electron gyromagnetic-ratio. The opposite precession rates drive the system away from a spin-temperature distribution due to differential rotation of the two hyperfine manifolds. For example, a spin oriented initially transverse to the $B_z$ field and prepared in a spin-temperature distribution $\rho\propto e^{\beta m_x}$ (with the quantization axis along $x$) will evolve after a precession time of $\pi/(2\gamma B_z)$ into a distribution $\rho\propto e^{(-1)^{\chi}\beta m_y}$, with the quantization axis along $y$, where $\chi=F-I-S$, as shown in Fig.~\ref{fig:energy_level}d. The dependence of the resulting distribution on the quantum number $F$ deviates from STD; collisions drive the spin degrees of freedom back toward STD, causing decoherence associated with spin-exchange collisions. For spin-ensembles aligned along the field, or when the Larmor frequencies are synchronized by optical means \cite{Berrebi2022}, there is no precession and therefore no spin-exchange decoherence. In the SERF regime, the deviation from STD is small; either the lower hyperfine manifold is barely populated at near unity spin-polarization, or the magnetic field perturbing the STD is weaker with respect to the spin-exchange collisions rate $\gamma B\ll R$.

\begin{figure*}[t]
\begin{centering}
\includegraphics[width=17.5cm]{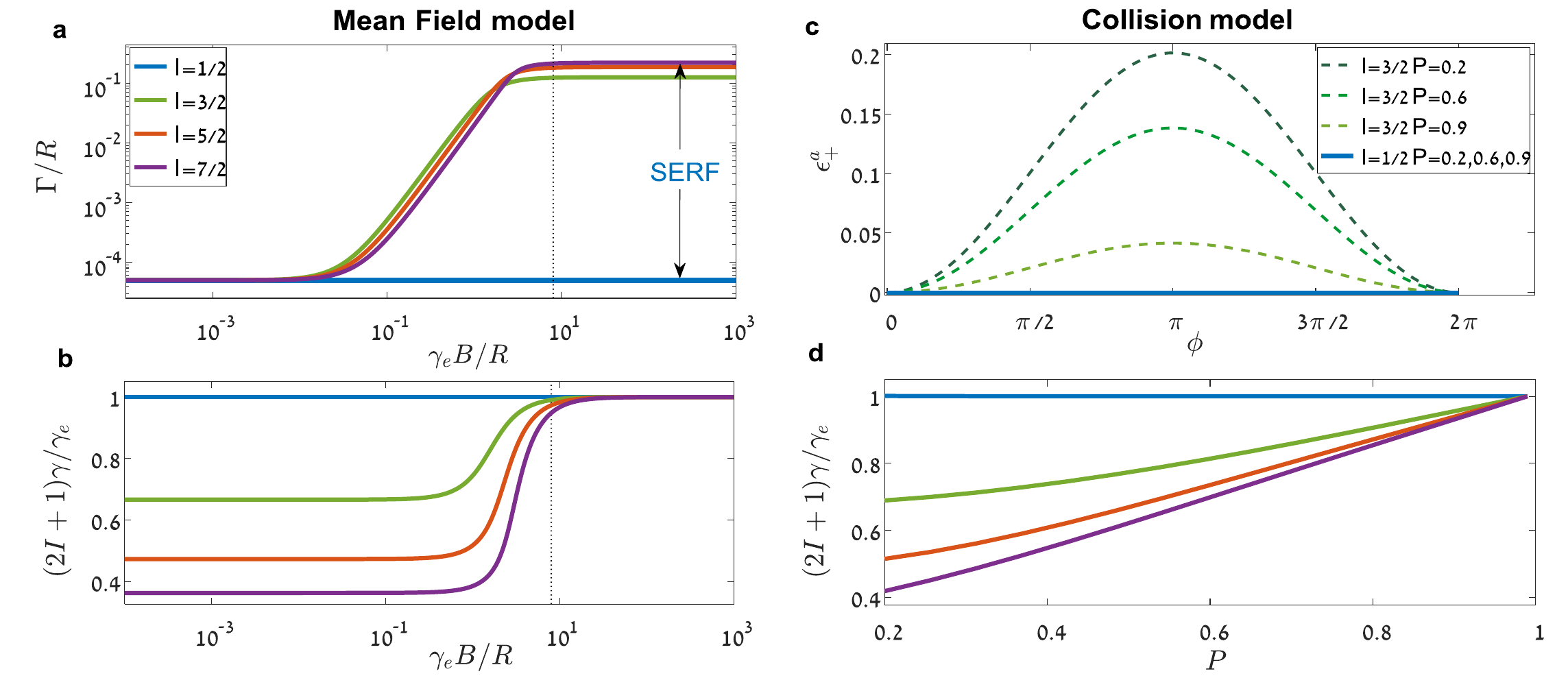}
\par\end{centering}
\centering{}\caption{\textbf{Suppression of Spin-Exchange relaxation and frequency slowing-down for $I=1/2$ atoms.} \textbf{a}, The transverse decoherence rate $\Gamma$ of $I>1/2$ atoms is dominated by spin-exchange at high magnetic field $B$. In contrast, the relaxation rate of $I=1/2$ atoms (blue) is highly suppressed at all magnetic fields; it is determined by the finite lifetime but independent of the spin-exchange rate $R$ .\textbf{b}, The gyromagnetic-ratio $\gamma$ depends on magnetic field magnitude for all atoms but $I=1/2$. \textbf{c}, Single events of spin-exchange collisions with relative scattering phase $\phi$ between singlet and triplet states redistribute magnetic coherence for all $I=3/2$ atoms in spin-temperature distribution, mostly at low degrees of polarization $P$, but not for $I=1/2$ atoms (at any $P$ in spin-temperature distribution). The unitless parameter $\epsilon$ quantifies the distribution of the magnetic coherence between the hyperfine manifold by collisions (See text). Since $\phi$ is random for such collisions in a thermal gas of atoms, the exchange process typically leads to decoherence. The nulling of $\epsilon$ for spin $I=1/2$ (blue line) for different degrees of spin polarization ($P=0.2$, $P=0.6$, and $P=0.9$ shown) highlights the suppression of spin-exchange relaxation for any $\phi$ and $P$. \textbf{d}, In the weak field limit ($\gamma_{\textrm{e}}B\ll R$), the gyromagnatic-ratio of $I>1/2$ atoms depends on the degree of spin polarization, but is constant for $I=1/2$ atoms. The gyromagnatic-ratio determines the slowing-down factor $q=\gamma_{\textrm{e}}/\gamma$. \textbf{a}-\textbf{b} are obtained from the hyperfine Bloch equations in the low polarization regime and \textbf{c}-\textbf{d} from a numerical stochastic model. \label{fig:SERF}}
\end{figure*}

For atoms with $I=1/2$ however, the lower hyperfine manifold is composed of a single level with $m=0$ as shown in Fig.~\ref{fig:energy_level}e-f. Consequently, a magnetic field generates precession of the magnetic moment in the upper hyperfine manifold only, and the STD is always maintained. It is therefore expected that atoms with $I=1/2$ would be inherently in the SERF regime, regardless of the strength of the Larmor precession $\gamma B$ and polarization $P$. 
\begin{figure*}[h]
\begin{centering}
\includegraphics[width=14.5cm]{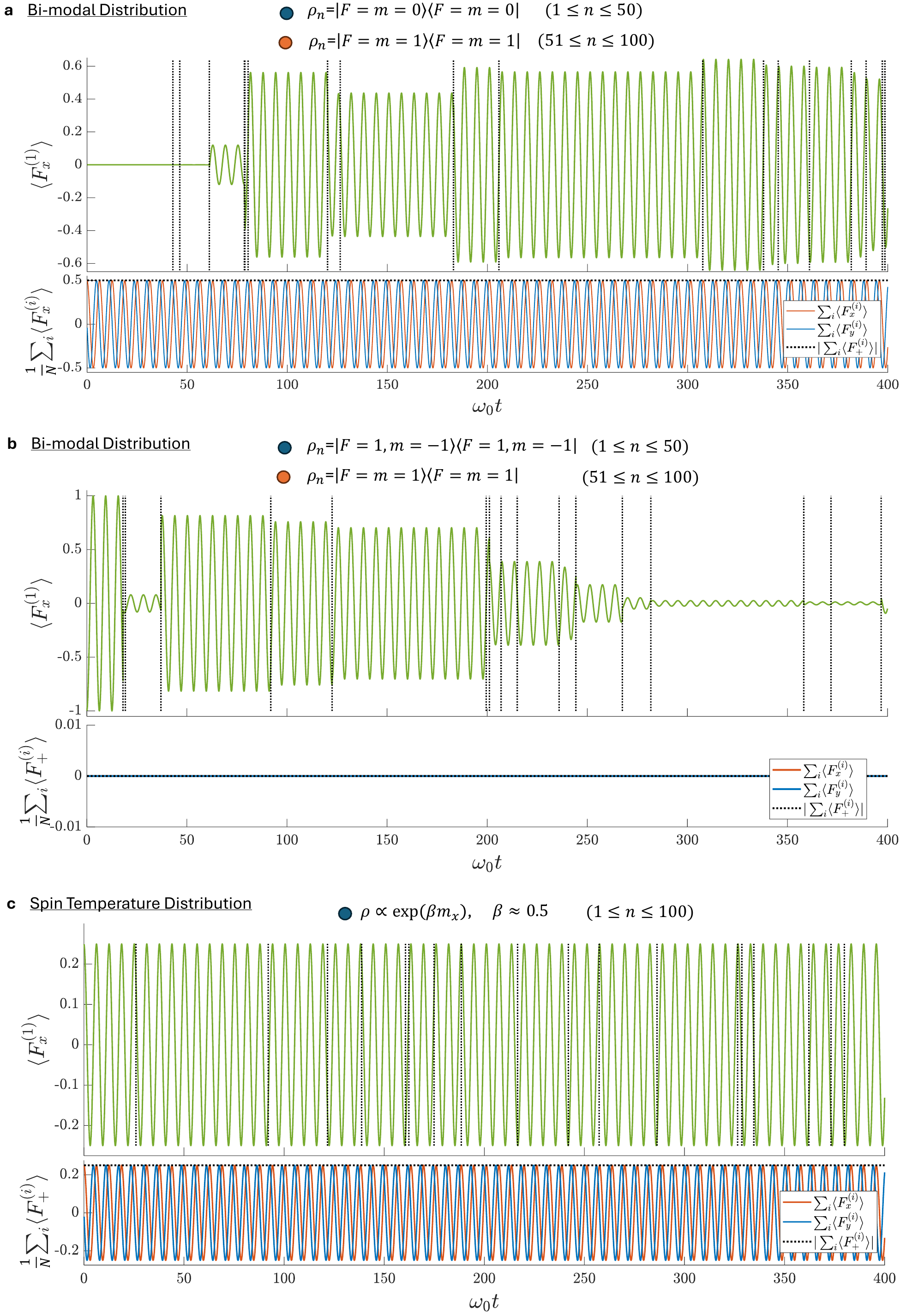}
\par\end{centering}
\centering{}\caption{\textbf{Many-body evolution of $S=I=\tfrac{1}{2}$ particles.} Stochastic evolution of $N=100$ particles under the Zeeman and Hyperfine Hamiltonian (Eq.~\ref{eq:H0}), with random spin-exchange events occurring at a mean rate of $1/R$ with $R=\omega/40$. At random collision times (vertical black lines), pairs of particles evolve according to Eq.~(\ref{eq:exchange_operator}) with a random $\phi$. Inter-atomic correlations are erased to maintain computational feasibility (see text). \textbf{a}, Evolution of an initial bi-modal distribution with half the particles in $\ket{F=m=1}$ and the other half in $\ket{F=m=0}$. \textbf{b}, Evolution of a bi-modal distribution with half the particles in $\ket{F=m=1}$ and the other half in $\ket{F=1,m=-1}$. \textbf{c}, Evolution of all particles initialized in a spin-temperature distribution with $P=1/2$. In all cases, the total spin of the ensemble $|\sum_i\langle F^{(i)}_{+}\rangle|$ (bottom plot) is conserved. In \textbf{a} and \textbf{b}, individual spins thermalize to the total spin value, while in \textbf{c}, the evolution of an individual spin remains stationary.} \label{fig:manybody_simulation}
\end{figure*}

We demonstrate this inherent suppression of spin-exchange relaxation for $I=1/2$ for spins initially in a STD, using two different models. First, we solve the hyperfine Bloch equations for atoms with nuclear spin $I$, and describe the dynamics of the average spin of the two hyperfine manifolds in the low polarization limit \cite{happer1977effect,katz2015coherent,Berrebi2022}. These equations account for the magnetic field precession at a rate $\gamma B$, spin exchange collisions at a rate $R=10^6\,\textrm{s}^{-1}$, and taking a fixed longitudinal spin lifetime $T_1=10\,\mathrm{ms}$ for all values of I. In Fig.~\ref{fig:SERF}(a-b) we present the decoherence rate $\Gamma$ for spins oriented transverse to the field and gyromagnetic ratio $\gamma_{\mathrm{eff}}$  as a function of magnetic field. For $I>1/2$ atoms, at high magnetic fields ($\gamma B\gg R$, practically including also geomagnetic fields, denoted by a dashed vertical line), the relaxation is dominated by spin-exchange collisions and is suppressed only at low magnetic fields. $I=1/2$ atoms in contrast, feature a constant and reduced decoherence rate which is solely governed by their finite lifetime. The gyro-magnetic ratio for $I=1/2$ atoms is also independent of the magnetic field, unlike all other atoms experiencing spin-exchange collisions. 

To further validate $I=1/2$ atoms' SERF property, we explore another model. We solve the Unitary spin-exchange evolution for pairs of alkali-like spins initially in a STD. During a collision, the two valence electrons of a pair decompose into Singlet and Triplet states, which acquire different phases, where their relative phase $\phi$ governs the transition amplitudes of the evolution. At room-temperature and above, $\phi$ is almost uniformly distributed between $0$ to $2\pi$ \cite{jau2010optically,pritchard1970alkali, burke1997impact, pritchard1967atomic, kartoshkin2010complex}. The outcome of a collision for an initial two-body density matrix $\rho_{a}\otimes\rho_{b}$ is then given by \cite{jau2010optically,katz2018synchronization}
\begin{equation}\label{eq:exchange_operator}\rho=\Pi_{\mathrm{T}}\rho_{a}\otimes\rho_{b}\Pi_{\mathrm{T}}+e^{i\phi} \Pi_{\mathrm{S}} \rho_{a}\otimes\rho_{b} \Pi_{\mathrm{S}},	\end{equation} where $\Pi_{\mathrm{S}}$  and $\Pi_{\mathrm{T}}$ are the projection operators on the singlet and triplet states respectively. We estimate the average amount of spin transfer between the two hyperfine states by the unitless quantity $\epsilon_+^{a}=(\langle F_{\mathrm{out}}^{(+a)}\rangle-\langle F_{\mathrm{in}}^{(+a)} \rangle)/(\langle F_{\mathrm{in}}^{(+a)} \rangle +\langle F_{\mathrm{in}}^{(+b)} \rangle)$ which is presented in Fig.~\ref{fig:SERF}c., using $\langle F_\mathrm{(in/out)}^{(+a)}\rangle =\textrm{Tr}(\rho F_+^{(a)} )$.  For $I=3/2$ atoms, $\epsilon_+^{a}>0$ and the magnetic moments are redistributed by collisions, depending on the values of $\phi$. The suppression of spin transfer parameter for $I>1/2$ is visible only at high degree of spin polarization, where $\epsilon_+^{a}(P\rightarrow1)\rightarrow0$. The magnetic moment in the upper hyperfine manifold for $I=1/2$ atoms, in contrast, remains unaffected by the collision for any $\phi$ and any $P$, demonstrating that for a spin-temperature distribution, the spin $\langle F_{\mathrm{out}}^{(+a)}\rangle=\langle F_{\mathrm{in}}^{(+a)}\rangle$ is unchanged by collisions.

Our results align with earlier studies that, while not specifically addressing the $I=1/2$ case, established general formulas for spin-exchange relaxation that are applicable to $I=1/2$. In the high polarization limit ($P\rightarrow 1$), spin-exchange relaxation is completely eliminated for mono-isotopic gases (e.g., see Eq.~168 in Ref.~\cite{appelt1998theory}). In the low-polarization limit ($P\ll1$), the numerical results in Fig.~\ref{fig:SERF}(a-b) agree with the analytical formulae in Ref.~\cite{happer1977effect} (Eqs.~101,102,106 and 107) for all values of $I$ in the low- and high-magnetic-field regimes, with $T_1\rightarrow\infty$. Furthermore, our results are consistent with the statistical analysis in  Ref.~\cite{happer1977effect} which attributes relaxation to collision-mediated jumps between different hyperfine levels. However, for $I=1/2$ atoms in a STD, the mean precession time in the upper hyperfine manifold before interruption approaches infinity (Eqs.~124 and 131 therein). This eliminates the estimated relaxation due to spin exchange (Eq.~134) and indicates that such jumps are suppressed, consistent with our derivation of $\epsilon_+^{a}=0$.


\section{Many-Body Simulations Beyond Spin-Temperature Distribution}
To this point, our results have focused on solutions where both colliding atoms follow a spin-temperature distribution, which is often a valid approximation under common experimental conditions \cite{anderson1959n, happer1987optical, appelt1998theory}. To explore scenarios beyond this assumption and examine the outcomes of collisions between atoms in different spin states, we extend our model to describe the many-body dynamics, adapting the simulation tools from Ref.~\cite{katz2018synchronization}. We construct a computational space of $N=100$ particles, each with nuclear spin $I=1/2$ and electron spin $S=1/2$, and evolve their state under the Hamiltonian: \begin{equation}\label{eq:H0}H_0=\hbar\sum_{n=1}^{N}2\omega_0 S_z^{(n)} + A_{\textrm{hpf}}\mathbf{I}^{(n)}\cdot\mathbf{S}^{(n)},\end{equation} which represents the Zeeman and hyperfine interactions, respectively. Here, $\mathbf{S}^{(n)}$ and $\mathbf{I}^{(n)}$ denote the vector spin operators of the spin-$\tfrac{1}{2}$ electron and nucleus of the $n$th particle, and $\omega_0=\gamma B_z$. The many-body state of the ensemble is initially given by $\rho=\rho_1\otimes\rho_2\otimes\cdots\otimes\rho_{N}$. For each particle, we assume that random spin-exchange collisions occur at a set of times $\{t_1,t_2,\ldots\}_n$, drawn from an exponential distribution with a mean time of $R^{-1}$. At each collision event, another particle in the ensemble is randomly selected, along with a random value of $\phi$ from a uniform distribution. The quantum states of the selected pair (with indices $a$ and $b$) are then instantaneously evolved according to Eq.~\ref{eq:exchange_operator}. To keep the computational space tractable, we  erase the inter-atomic coherence developed after each collision event by performing a partial trace, thereby maintaining the form  $\rho=\rho_1\otimes\rho_2\otimes\cdots\otimes\rho_{N}$.

In Fig.~\ref{fig:manybody_simulation}a (top) we present the stochastic dynamics of a single particle for $\omega_0=40R$ during a single simulation run, assuming an initially bi-modal spin distribution. Half of the particles ($1\leq n\leq50$) are initialized in the lower hyperfine manifold $\rho_n=\ket{F=m=0}\bra{F=m=0}$, while the other half ($51\leq n\leq100$) are in the maximally polarized spin state, $\rho_n=\ket{F=m=1}\bra{F=m=1}$, with the quantization axis along $x$. Collision events with other particles in the ensemble are marked by dotted vertical lines. Evidently, several collisions result in discontinuous amplitude changes, with the evolution eventually approaching a stationary precession state that tracks the mean spin of the ensemble.

Remarkably, the total spin of the ensemble, \begin{equation}|\textbf{F}_+|=|\sum_n\langle F^{(n)}_+\rangle|=\sqrt{|\sum_n\langle F^{(n)}_x\rangle|^2+|\sum_n\langle F^{(n)}_y\rangle|^2},\end{equation} is conserved and exhibits no stochastic evolution, as shown in Fig.~\ref{fig:manybody_simulation}a (bottom). We repeat the simulation for a different initial bi-modal distribution, where particles $1\leq n\leq50$ are initialized in $\rho_n=\ket{F=1,m=-1}\bra{F=1,m=-1}$. The resulting evolution is shown in Fig.~\ref{fig:manybody_simulation}b. In this case, the individual particle thermalizes toward an unpolarized state (a spin-temperature distribution with $P=0$), while the mean spin is conserved at zero.

Finally, in Fig.~\ref{fig:manybody_simulation}c, we present the evolution of all particles initialized in a spin-temperature distribution with $P=0.25$ (corresponding to $\beta\approx 0.5$) using $x$ as the quantization axis. The results demonstrate no jumps, consistent with our calculation of $\epsilon_+^{a}=0$. Using these simulations, we also characterize the gyromagnetic ratio as a function of the spin polarization $P$ in the low magnetic field regime, with the results presented in Fig.~\ref{fig:SERF}d. Interestingly, unlike $I>1/2$ atoms, the gyromagnetic ratio of $I=1/2$ atoms is independent of $P$.

The conservation of the total spin beyond the spin-temperature distribution (STD), as observed in Fig.~\ref{fig:manybody_simulation}{a-b}, can be explained using straightforward arguments. The total spin operator of the ensemble is invariant under exchange evolution because the total electron and nuclear spins of colliding pairs are conserved. Additionally, the total spin operator commutes with the single-particle hyperfine interaction. For $I=\tfrac{1}{2}$ and  within the regime of the linear Zeeman effect ($\omega_0\ll A_{\textrm{hpf}}$), the evolution of the total spin in the Heisenberg picture is given by $\textbf{F}_+(t)=e^{i\omega_0t}\textbf{F}_+(0)$ ensuring that $|\langle \textbf{F}_+\rangle|$ remains conserved for any quantum state. In contrast, for $I>\tfrac{1}{2}$, Larmor precession in opposite directions within the two hyperfine manifolds can modulate $|\langle \textbf{F}_+\rangle|$. When combined with spin-exchange interactions, this modulation can lead to a decay of the total spin. We note that the results and discussion in this work are limited to Zeeman coherences, whereas hyperfine coherences, which are utilized in masers and clocks \cite{kleppner1965hydrogen}, may decay rapidly by spin-exchange collisions \cite{happer1977effect}.

\section{Hydrogen Magnetometer}
The hydrogen atom has a single valence electron and its nuclear spin is $I=1/2$. It could therefore be utilized as an inherent SERF magnetometer. Its most stable form appears as a molecule $\mathrm{H}_2$, which can be efficiently dissociated via e.g., application of high-frequency RF discharge. Once dissociated, it can maintain its atomic form for long time before associating back into a molecule as this process typically requires three bodies \cite{slevin1981radio,longo1998nonequilibrium,xiao2016charge,goodyear1962dissociation,bell1972model,macdonald1949high,rose1955high}; $1$ Torr of atomic hydrogen in a cm-size enclosure can last about a hundred of milliseconds, as traditionally utilized in precision masers \cite{busca2003cronos,zhang2001study,kleppner1965hydrogen,goldenberg1960atomic,peters1987hydrogen}.

A primary challenge in usage of atomic hydrogen relates to its energetic excited state, which hinders optical manipulation or readout of its ground state using standard laser frequencies, since its optical transitions are in the UV. Masers overcome this challenge by using non-optical methods \cite{ruff1965spin,walsworth1992measurement,humphrey2003testing}; pumping the hydrogen spins using permanent magnets in a Stern-Gerlach configuration, and reading their state using microwave cavities. Here we consider another approach based on a hybrid configuration where the hydrogen is collisionally coupled to another optically accessible ensemble of alkali-metal atoms \cite{poelker1994high,cole1985spin,redsun1990production,anderson1995spin}. 

We consider a gaseous mixture of atomic hydrogen with potassium atoms ($I=3/2)$ as illustrated in Fig.~\ref{fig:DUAL_MAGNETOMETER}a. The potassium spin-polarization vector $\textbf{P}_\mathrm{K}$ is optically-pumped continuously  along the magnetic field $B_z\hat{z}$, and its transverse component in the $xy$ plane can be optically probed using standard techniques \cite{katz2021coupling,mazzinghi2021cavity}. The potassium spins are coupled to the ensemble of hydrogen spins with mean polarization vector $\textbf{P}_\mathrm{H}$ via mutual spin exchange collisions, which act to equilibrate the spin-polarization of the two ensembles {\cite{dehmelt1958spin}.

\begin{figure}[t]
\begin{centering}
\includegraphics[width=8.4cm]{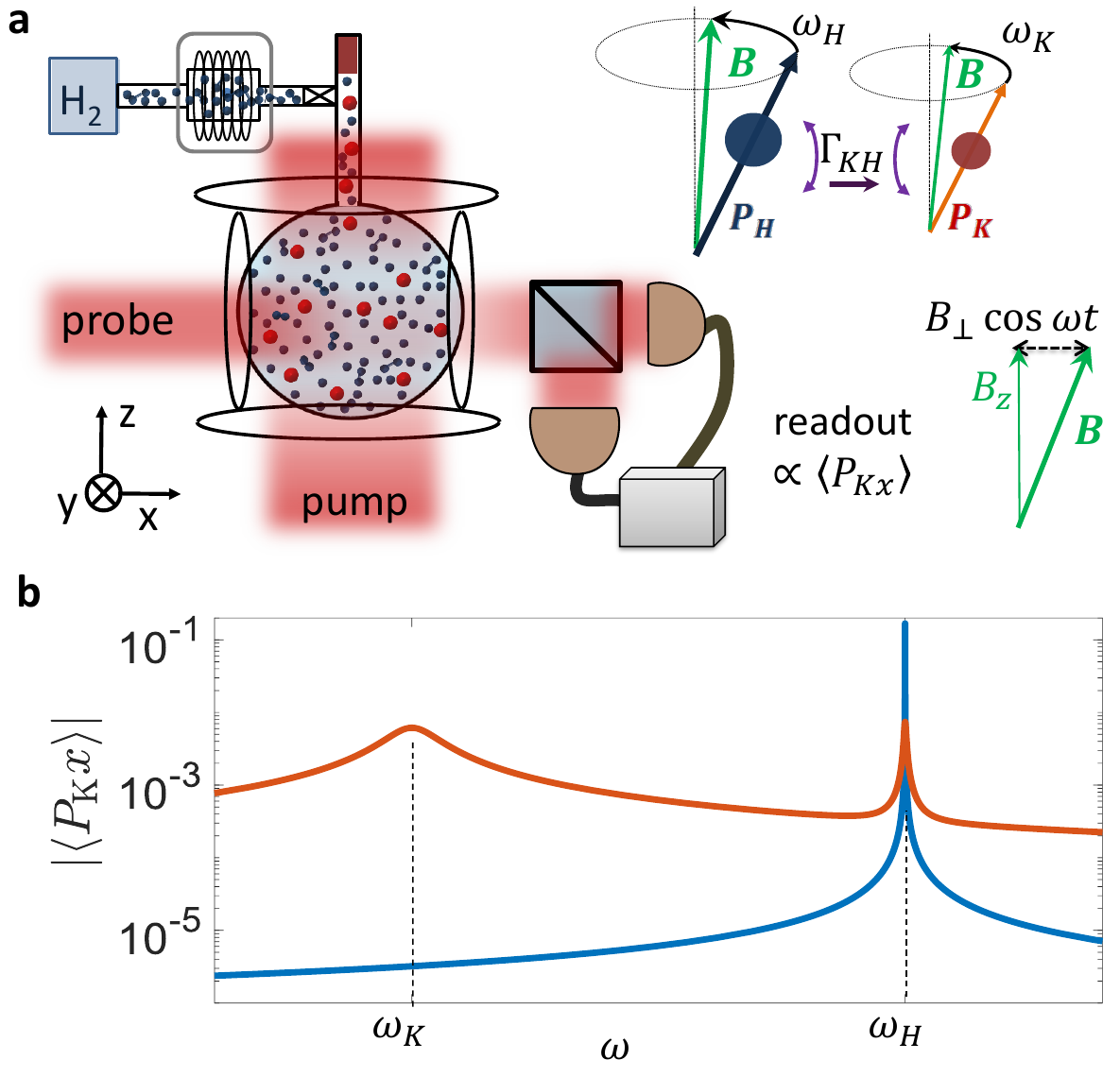}
\par\end{centering}
\centering{}\caption{\textbf{Operation of a dual potassium-hydrogen magnetometer.} \textbf{a}, Configuration of a dual-specie magnetometer. Atomic hydrogen ($I=1/2$) precesses around a magnetic field $B_z\hat{z}$ using a weak magnetic drive $B_{\perp}\cos(\omega t)\hat{x}$, and imprints its precession on a dilute potassium spin gas ($I=3/2$). The optically-controlled potassium spins continuously polarize and read the state of the hydrogen spins. \textbf{b}, Calculated magnetic response of the potassium spins to the oscillatory magnetic drive. The potassium spins, dressed by the interaction with the hydrogen spins, feature a double resonance spectrum with resonance frequencies $\omega_{\textrm{K}}$ and $\omega_{\mathrm{H}}=2\omega_{\mathrm{K}}$, as shown in red line for a low hydrogen density. At optimal conditions, the hydrogen density is higher and the imprinted hydrogen response dominates the potassium magnetic spectrum (blue line). Red and blue lines correspond to cross and star symbols in Fig.~\ref{fig:optimal}, respectively. \label{fig:DUAL_MAGNETOMETER}}
\end{figure}

This collisional coupling enables pumping of the hydrogen spins and also imprinting their precession on the measured response of the potassium. We can force the hydrogen precession by application of a weak and transverse magnetic field $B_\perp \cos(\omega t)\hat{x}$, whose modulation is resonant with the hydrogen frequency $\omega_\mathrm{H}=\gamma_{\mathrm{H}}B$ but off resonant from the potassium $\omega_\mathrm{K}=\gamma_{\mathrm{K}}B=\omega_\mathrm{H}/2$ at high magnetic fields. 
Including spin-relaxation of the valence electron for both species at rates $R_{\mathrm{sd}}^\mathrm{K}$, $R_{\mathrm{sd}} ^\mathrm{H}$ and optical pumping of the potassium electrons at rate $R_\mathrm{p}$ along $\hat{z}$, we can describe the coupled dynamics with the set of Bloch equations\begin{equation}
\begin{aligned}
\dot{\textbf{P}}_\mathrm{H}&=\gamma_\mathrm{H} \textbf{B}\times \textbf{P}_\mathrm{H}+\Gamma_{\mathrm{HK}}\textbf{P}_\mathrm{K}-\Gamma_\mathrm{H} \textbf{P}_\mathrm{H}+\boldsymbol{\xi}_\mathrm{H},\\
\dot{\textbf{P}}_\mathrm{K}&=\gamma_{\mathrm{K}}\textbf{B}\times \textbf{P}_{\mathrm{K}}+\Gamma_{\mathrm{KH}}\textbf{P}_{\mathrm{H}}-\Gamma_{\mathrm{K}}\textbf{P}_{\mathrm{K}}+\Gamma_{\mathrm{p}}\hat{z}+\boldsymbol{\xi}_\mathrm{K}.
\end{aligned}
\label{eq:Bloch-dynamics}
\end{equation} Here $\textbf{B}(t)=B_z\hat{z}+B_{\perp}\cos(\omega t)\hat{x}$ denotes the total magnetic field vector and $\Gamma_{\mathrm{HK}}=k_{\mathrm{HK}}n_\mathrm{K}/q_{\mathrm{H}}$ and $\Gamma_{\mathrm{KH}}=k_{\mathrm{HK}}n_\mathrm{H}/q_{\mathrm{K}}$ are the hybrid spin-exchange rates, where $n_{\mathrm{K}}$,$n_{\mathrm{H}}$ are the potassium and hydrogen number-densities, $k_{\mathrm{HK}}=5.4\times10^{-10} ~\mathrm{cm}^3/\mathrm{s} $ is the mutual spin-exchange rate coefficient \cite{cole1985spin} and $q_{\mathrm{H}}$ and $q_{\mathrm{K}}$ are the slowing-down factors of the two species \cite{appelt1998theory} presented in Fig.~\ref{fig:SERF}d. We take  $\Gamma_{\mathrm{p}}=R_\mathrm{P}/{q}_\mathrm{K}$ as the optical pumping rate and $\Gamma_\mathrm{K}=\Gamma_{\mathrm{p}}+\Gamma_{\mathrm{KH}}+(R_{\mathrm{sd}}^\mathrm{K}+R_{\mathrm{se}}^\mathrm{K})/{q}_{\mathrm{K}}$ and $\Gamma_{\mathrm{H}}=\Gamma_{\mathrm{HK}}+R_{\mathrm{sd}}^\mathrm{H}/{q}_\mathrm{H}$ as the transverse relaxation rates of the two ensembles. The hydrogen spins relax by spin-destruction processes or spin-exchange collisions with the potassium spins, but crucially, are free of spin-exchange relaxation by rapid collisions with other hydrogen atoms (i.e.,~$R_{\mathrm{se}}^\mathrm{H}=0$).

We introduce the white noise vector processes $\boldsymbol{\xi}_{\mathrm{H}}$ and $\boldsymbol{\xi}_{\mathrm{K}}$ transverse to the polarization axis to describe the Atom Projection Noise (APN) that limits the fundamental sensitivity of the magnetometer (i.e.~the standard quantum limit) \cite{dellis2014spin,mouloudakis2022effects}. They satisfy $\langle\boldsymbol{\xi}_{\mathrm{H}}(t)\rangle=\langle \boldsymbol{\xi}_{\mathrm{A}}(T)\rangle=0$ but have a nonzero variance $\langle \xi_{\mathrm{qi}}(t)\xi_{\mathrm{q'j}} (t')\rangle=\delta_{ij}\delta_{qq'}\delta(t-t')R_q/(n_qV)$ for $q,q'\in\{\mathrm{K,H}\}$ and $i,j\in\{x,y\}$. 

We begin by presenting an approximate analytical solution of Eqs.~(\ref{eq:Bloch-dynamics}) to illustrate the coupled dynamics, and then present more general numerical results. We consider the regime in which the Hydrogen is the dominant specie $n_{\mathrm{H}}\gg n_{\mathrm{K}}$ and assume the rotating wave approximation such that $\omega\gtrsim \Gamma_{\mathrm{K}},\Gamma_{\mathrm{KH}}\gg \Gamma_{\mathrm{H}}$. By continuous optical pumping the potassium spins reach a steady polarization of \begin{equation}P_{\mathrm{K}z}=\frac{\Gamma_\mathrm{p}\Gamma_{\mathrm{H}}} {\Gamma_{\mathrm{K}}\Gamma_{\mathrm{H}}-\Gamma_{\mathrm{KH}}\Gamma_{\mathrm{HK}}}.\end{equation}
Sizeable polarization requires pumping rate that scales as  $\Gamma_{\mathrm{p}}\sim n_{\mathrm{H}}k_{\mathrm{HK}},$ accounting for the transfer of spin polarization from the rare potassium vapor to the dense hydrogen gas. In the presence of the weak driving field $\gamma_{\mathrm{H}}B_{\perp}\lesssim \Gamma_{\mathrm{H}},$ the mean response of the transverse potassium spin $\langle P_{\mathrm{K}+} \rangle=\langle P_{\mathrm{K}x}+iP_{\mathrm{K}y}\rangle$ at the drive frequency $\omega$ is dominated by the coupling to the driven hydrogen and in the rotating frame is given by \begin{figure}[t]
\begin{centering}
\includegraphics[width=8.6cm]{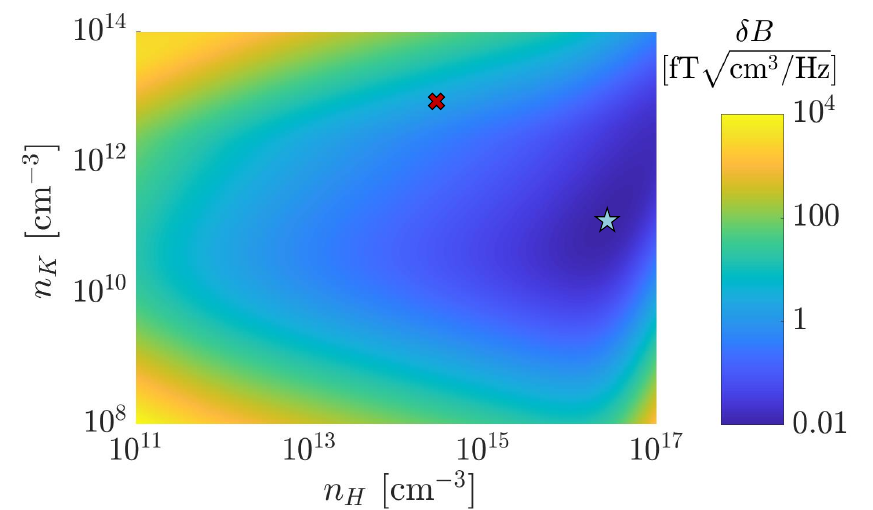}
\par\end{centering}
\centering{}\caption{\textbf{Projected Fundamental sensitivity.} The magnetic sensitivity of the hydrogen-potassium magnetometer limited by atom-projection noise, and independent of the magnetic field magnitude. Cross and star correspond to the configurations presented in Fig.~\ref{fig:DUAL_MAGNETOMETER}b in red and blue lines respectively. \label{fig:optimal}}
\end{figure}
\begin{align} \label{eq:nested_resonance}  
\langle P_\mathrm{K+}(\omega)\rangle=\frac{i\gamma_\mathrm{K}B_{\perp}P_\mathrm{Kz}}{\Gamma_{\mathrm{H}}(1-\frac{(\Gamma_\mathrm{K}-i(\omega_{\mathrm{K}}-\omega))}{\Gamma_{\mathrm{KH}}}\frac{(\Gamma_\mathrm{H}-i(\omega_{\mathrm{H}}-\omega))}{\Gamma_{\mathrm{HK}}})}.
\end{align}
The spectral response of Eq.~(\ref{eq:nested_resonance}) has a double resonance shape, a broad complex Lorentzian centered at $\omega_{\mathrm{K}}$ with linewidth $\Gamma_{\mathrm{K}}$, and a narrow complex Lorentzian centered at $\omega_{\mathrm{H}}$ and linewidth $\Gamma_{\mathrm{H}}$, as shown in Fig.~\ref{fig:DUAL_MAGNETOMETER}b (red line). The former corresponds to the response of the potassium to oscillations at its bare resonance, which is broadened by rapid collisions with the hydrogen and optical pumping. The latter and stronger resonance, manifests the efficient response of hydrogen atoms to the field, imprinted on the potassium response by rapid spin-exchange collisions. 

The narrow resonance line at $\omega=\omega_{\mathrm{H}}$ is associated with a sharp variation of the phase of $\langle P_{\mathrm{K}+} \rangle$ which in turn allows for sensitive estimation of the resonance frequency $\omega_{\mathrm{H}}$ and therefore of the magnetic field $B_z$. We estimate the magnetic sensitivity in this configuration due to APN by linearizing the response of Eq.~(\ref{eq:nested_resonance}) around $\omega-\omega_{\mathrm{H}}$ and computing the APN variance, see Appendix A. To reveal the qualitative scaling of the magnetic noise, we assume the simplifying conditions $\Gamma_{\mathrm{H}}\sim \Gamma_{\mathrm{HK}}$ and $\Gamma_{\mathrm{K}}\sim \Gamma_{\mathrm{KH}}$ and find that for a measurement volume $V$, the spin variance of the potassium scales as $\langle|\Delta P_{\mathrm{K}+}(\omega=\omega_{\mathrm{H}})|^2\rangle\propto t/(\Gamma_{\textrm{H}}n_{\mathrm{H}}V)$, and the fundamental magnetic sensitivity is limited by the effective magnetic noise
\begin{equation} 
\delta B_{\textrm{APN}}\propto\frac{1}{\gamma_{\mathrm{H}}}\sqrt{\frac{\Gamma_{\textrm{H}}}{n_\mathrm{H}V}}. 
\label{eq:FMN}
\end{equation}
Remarkably, the fundamental sensitivity in this configuration is determined by the Hydrogen properties: its high number-density and its narrow linewidth, therefore enabling high sensitivity at large magnetic fields $\gamma_\mathrm{H} B\gg \Gamma_\mathrm{H}$. Comparison of this response to a single-specie potassium magnetometer, whose low density is tuned to set $\Gamma_{\textrm{K}}=\Gamma_{\textrm{H}}$, reveals that the sensitivity in the proposed hybrid configuration is enhanced by a large factor that scales as $\sqrt{n_\mathrm{H}/n_\mathrm{K}}$.

\section{Numerical Optimization of Fundamental Sensitivity}
To quantitatively analyze the magnetometer performance, we numerically calculate the fundamental sensitivity as a function of $n_\mathrm{K}$ and $n_\mathrm{H}$, and at each point optimize for the optical pumping rate $\Gamma_\mathrm{p}$ and the amplitude of the magnetic drive $B_{\perp}$. We take $\Gamma_{\mathrm{H}}=40\,\textrm{s}^{-1}$ and present the calculated fundamental sensitivity in Fig.~\ref{fig:optimal}. We find a fundamental magnetic sensitivity of $\delta B_{\textrm{APN}} =10  \textrm{aT} \sqrt{\mathrm{cm}^{3}/\textrm{Hz}}$ as marked by a cyan star for $n_\mathrm{K}=1.2\times10^{11}\,\textrm{cm}^{-3}$, $n_\mathrm{H}=2.7\times10^{16}\,\textrm{cm}^{-3}$ using $B_{\perp}=0.35\,\textrm{nT}$ and $\Gamma_p=1.2\times10^{7}\,\textrm{s}^{-1}$, which correspond to about $1.5\,\mathrm{W}$ of pump beam power in a $1"$ diameter cell. The spectral response of the potassium at these conditions is plotted in Fig.~{\ref{fig:DUAL_MAGNETOMETER}} with a blue line and the configuration associated with the red line is marked with a red cross in Fig.~\ref{fig:optimal}.

Estimating the scale of technical noise mechanisms, specifically photon shot noise (PSN), is also interesting. This noise arises from fluctuations in the number of photons comprising the probe signal. Expressed in terms of magnetic field uncertainty, the noise is approximately given by \cite{smullin2009low} \begin{equation}\label{eq:PSN}\delta B_{\textrm{PSN}} = \frac{\sqrt{2}}{\sqrt{\Phi}\pi L n_\mathrm{K} r_e c f D(\Delta)}\left|\frac{\partial \langle P_{\mathrm{K}}^{+}\rangle}{\partial B_z}\right|^{-1}\end{equation}
where $\Phi$ is the photon flux integrated over the probe beam area, which is assumed to cover the entire cell; $f=0.34$ is the oscillator strength for the $\textrm{D}_1$ transition; $r_e=2.8\times10^{-13}$ cm is the classical electron radius; and $c$ is the speed of light. The term $D(\Delta)$ represents the dispersive part of the Voigt spectral line-shape \cite{happer2010optically}, accounting for pressure broadening, the natural lifetime, and Doppler broadening. For large optical detunings $\Delta$ from the optical transition, $D(\Delta)$ can be approximated as $D\approx\Delta^{-1}$. The parameter $L$ is the number of optical paths through the cell, which, for a single passage, is approximately the cell diameter. For the aforementioned computed configuration (blue star in Fig.~\ref{fig:optimal}), we find the magnetometer response to be  $|\partial\langle P_{\mathrm{K}}^{+}\rangle/{\partial B_z}|\approx4\times10^{8}\,\textrm{T}^{-1}$. Using a probe beam of 1 Watt power, detuned by $2$ GHz from the optical transition, we estimate $\delta B_{\textrm{PSN}}\approx 8 \textrm{aT}/\sqrt{\textrm{Hz}}$. Notably, the relaxation rate due to probe photon absorption in this configuration, $\Gamma_{\textrm{pr}}\approx 2\pi\times 1$ kHz, is very small, satisfying $\Gamma_{\textrm{pr}}\ll\Gamma_{\textrm{K}}$. This leaves significant potential for further improvement of $\delta B_{\textrm{PSN}}$, achievable through longer optical paths using multi-pass optics (e.g., \cite{herriott1965folded,sheng2013subfemtotesla}), smaller detunings, or higher optical power.

We now estimate the effect of other relaxation mechanisms in this configuration. For low buffer gas pressure, wall relaxation could be dominant unless anti-relaxation coatings are used. For paraffin-coated cells, it has been shown that hydrogen and alkali-metal atoms can bounce from the walls $\sim 10^4$ times before the spin state is appreciably perturbed. Assuming ballistic trajectories, we expect a decoherence rate of $\Gamma_{\textrm{H}}^{(\textrm{wall})}\approx 31\,\textrm{s}^{-1}$ and $\Gamma_{\textrm{K}}^{(\textrm{wall})}\approx 4\,\textrm{s}^{-1}$ for the proposed configuration. Magnetic field inhomogeneity is another source of decoherence, particularly when oriented along the static field. As the atoms move within the cell at thermal velocities and bounce from the walls, they also average the magnetic field they probe within the cell. We simulated this effect numerically for a linear magnetic field gradient, as detailed in Appendix \ref{dB_restrictions}, with the results shown in Fig.~\ref{fig:Bgrad_simulation}. For a geomagnetic field of $B=0.4\,\textrm{G}$, we expect that a gradient of about $\partial B_z/\partial z=0.69~\textrm{mG/cm}$ would lead to a decoherence rate $\Gamma_{\textrm{H}}^{(\nabla B)}=40\,\textrm{s}^{-1}$.

\section{Discussion}
In summary, we proposed and analyzed using the mixture of hydrogen-potassium atoms for a precision optical magnetometer. Owing to their simple spin structure, hydrogen atoms ($I=1/2$) are free of spin-exchange relaxation at any magnetic field and spin-polarization, enabling the potential operation of this magnetometer in the ultra-sensitive SERF regime at geomagnetic fields.

It is interesting to compare the operation of this hybrid magnetometer with other precision dual species sensors such as alkali-metal and noble gas comagnetometers \cite{gentile2017optically,vasilakis2009limits,shaham2022strong,katz2021coupling}. In comagnetometers, the alkali spin is utilized for pumping and probing of noble-gas spins, which are used for precision sensing of external fields. The coupling between noble-gas atoms to alkali-metal spins relies on the weak Fermi-contact interaction, and can attain high sensitivity which is limited to an ultranarrow bandwidth set by their long lifetime \cite{bloch2020axion,jiang2021search,bloch2022new}. Our configuration in contrast, relies on strong exchange interaction between the valence electrons of the two species, and the efficient response of the hydrogen atoms to magnetic fields.

Furthermore, the hybrid hydrogen-potassium configuration is potentially more accurate with respect to other single-specie SERF magnetometers; The hydrogen's gyromagnetic ratio is known to a high accuracy, and unlike atoms with $I>1/2$, it is independent of the degree of spin polarization or magnitude of the magnetic field. 

Finally, it is intriguing to consider the operation of this configuration from a spin-noise perspective. The spin noise of dual specie configurations has been analyzed theoretically and experimentally in \cite{dellis2014spin}, demonstrating that spin-exchange coupling with another specie acts to increase the measured noise variance. Here we also find that the potassium spin-projection noise is increased by a factor $\sim\sqrt{n_\mathrm{H}/n_\mathrm{K}}$. It is the improvement of the potassium signal by a larger factor  $\sim{n_\mathrm{H}/n_\mathrm{K}}$  that leads to the enhanced sensitivity of this hybrid sensor. The dressed hydrogen-potassium state  could also apply in  emerging avenues like efficient generation of spin-entanglement \cite{mouloudakis2021spin,kong2020measurement,guarrera2021spin,kominis2008sub,bao2020spin} and other quantum optics applications \cite{hammerer2010quantum,katz2020long,serafin2021nuclear,julsgaard2004experimental,katz2022optical}.

\clearpage
\newpage

\appendix 

\section{Hyperfine Bloch Equations}\label{Hyperfine_Bloch_Equations}

In this appendix we present the hyperfine-Bloch equations, which we used for estimation of the relaxation rate and the gyromagnetic ratio appearing in Fig. 2a,b in the main text. The average transverse spins at the two hyperfine manifolds  $a=I+ \frac{1}{2}$ and $b=I- \frac{1}{2}$ are given by ~(\ref{eq:SAEq1}).

Where $F_{a+}=F_{ax}+iF_{ay}$ and $F_{b+}=F_{bx}+iF_{by}$ are the transverse spin operators in a complex form. We solve these equations by finding the eigenvalues whose real part is associated with the relaxation and imaginary part with the precession frequency. We use the values given in  Table \ref{tab:coeff} and take $R_{\mathrm{SE}}=10^6 \mathrm{s}^{-1}$. Note that we scale $R_{\mathrm{SD}}(I)$ to maintain a constant lifetime of $T_1=10 {\mathrm{ms}}$. We solve those equations as a function of the magnetic field $B$ which enters through $\omega_{\mathrm{e}}=g_{\mathrm{e}}B$.

\begin{table}[h!]
\centering
\begin{tabular}{|c|c|c|c|c|}
 \hline
 $\boldsymbol{I}$ & $\boldsymbol{1/2}$ &  $\boldsymbol{3/2}$ &  $\boldsymbol{5/2}$ &  $\boldsymbol{7/2}$ \\ [0.25ex] 
 \hline \hline
$x_a$ & $1/2$ & $7/16$ &  $12/27$ &  $29/64$  \\ 
 \hline
$x_b$ & $1$ & $11/16$ &  $33/54$ &  $37/64$  \\ 
 \hline
$x_{ab}$ & $0$ & $-\sqrt{45}/16$ & $-\sqrt{630}/54$ & $-\sqrt{945}/64$  \\ 
 \hline
$y_a$ & $0$ & $1/8$  & $5/27$  & $7/32$  \\ 
 \hline
$y_b$ & $1$ & $5/8$ & $14/27$ & $15/32$  \\ 
 \hline
$y_{ab}$ & $~~~~~~~~~~~0~~~~~~~~~~~$ & $-\sqrt{5}/8$ & $-\sqrt{70}/27$ & $-\sqrt{105}/32$  \\ 
 \hline 
$a$ & $1$ & $2$ & $3$ &  $4$  \\  
\hline
$b$ & $0$ &  $1$ & $2$  &  $3$  \\  
\hline 
$R_{\mathrm{SD}}$ & $100$ & $300$ & $633$ & $ 1100 $  \\ 
 \hline
\end{tabular}
 \caption{Table I: Parameters used to solve  Eq.~(\ref{eq:SAEq1}) for different values of the nuclear spin $I$.}
\label{tab:coeff}
\end{table}

\begin{widetext}
\begin{align} \label{eq:SAEq1}
\frac{d}{dt}
\begin{pmatrix}
\left\langle{F_a}_+\right\rangle\\
\left\langle{F_b}_+\right\rangle
\end{pmatrix}
=\left(\frac{-i\omega_e}{2I+1}\begin{pmatrix}1&0\\0&-1\end{pmatrix}-
R_{SD}\begin{pmatrix}x_a&x_{ab}\\x_{ab}&x_b\end{pmatrix}-
R_{SE}\begin{pmatrix}y_a&y_{ab}\\y_{ab}&y_b\end{pmatrix}
\right)
\begin{pmatrix}
\left\langle{F_a}_+\right\rangle\\
\left\langle{F_b}_+\right\rangle
\end{pmatrix}
\end{align}

\end{widetext}

\section{Dual specie Spin dynamics}

We present the Bloch equations for the $\mathrm{H}-^{39}\mathrm{K}$ hybrid configuration, detail the exact procedure solved numerically and derive an approximated analytical form.

\begin{figure*} [htb]
\begin{centering} 
\includegraphics[width=17.0cm]{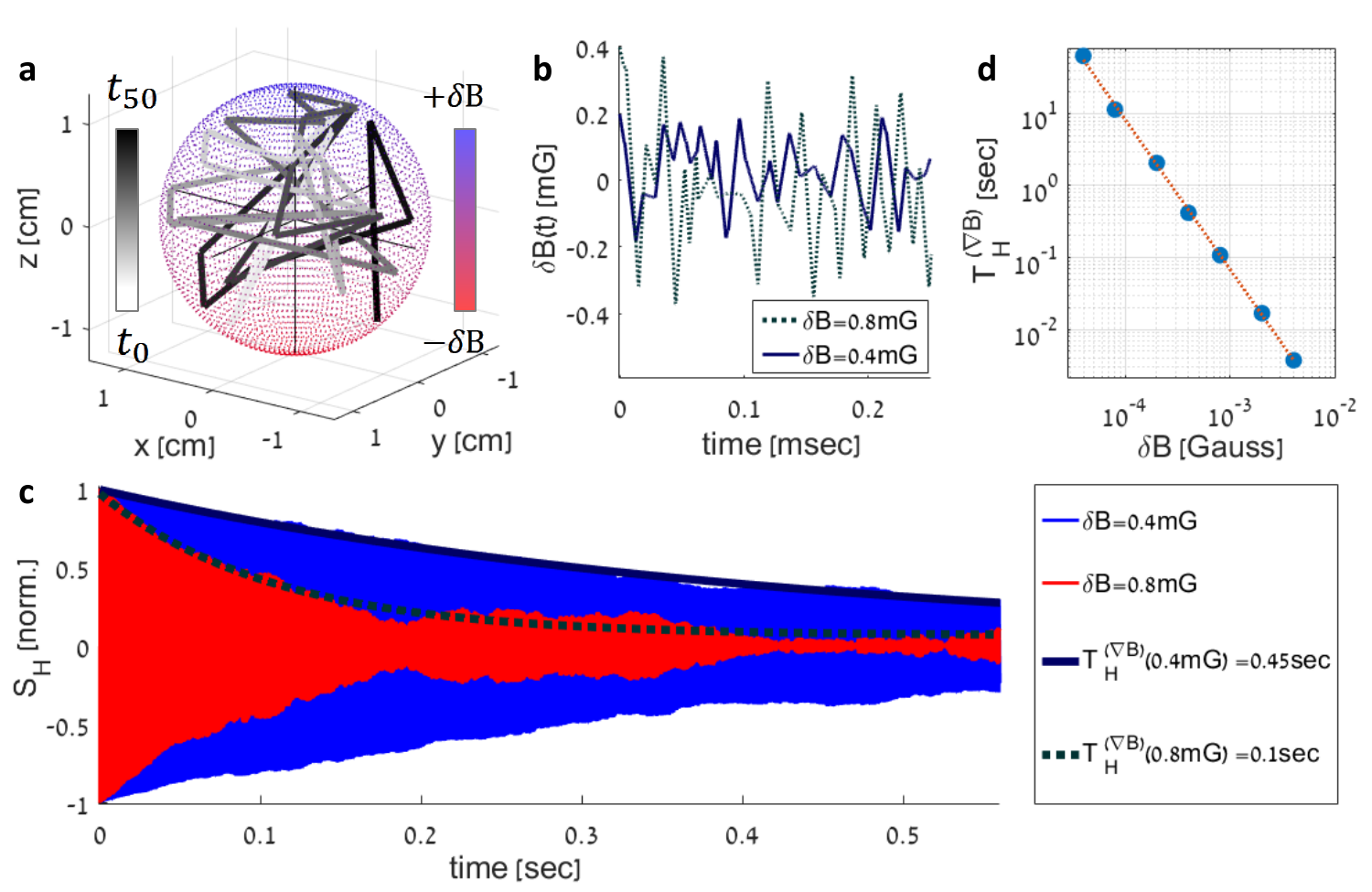}
\end{centering}
\centering{}\caption{\textbf{Decoherence by Magnetic Field Gradient}
\textbf{a}: Numerical simulation illustrating the ballistic trajectory of a hydrogen atom within the cell, with random reflection directions from the wall's normal. The red-blue colorbar indicates the strength of the local magnetic field (linear gradient along the $z$ axis), while the gray-black time scale represents time, showing the first 50 collisions.
\textbf{b}: Example demonstrating the magnetic field probed by an atom for two distinct trajectories and magnetic field strengths.
\textbf{c}: Ensemble average of the spin precession signal $S_{\textrm{H}}$ over time. Variations in the magnetic field within the cell lead to the observed exponential decay. Continuous and dashed lines represent fits to envelope decays for $\delta B = 0.4$ and $0.8$ mG, respectively.
\textbf{d}: Fitted coherence time $T_{\textrm{H}}^{(\nabla B)}$ as a function of gradien strength.\label{fig:Bgrad_simulation}}
\end{figure*}

\subsection{Approximated analytical solution derivation}

The equations of motion from the main text are given by 

\begin{align}
\dot{P}_{\mathrm{H}}&=\gamma_{\mathrm{H}}B\times P_{\mathrm{H}}+\mathrm{\Gamma}_{\mathrm{HK}}P_{\mathrm{K}}-\mathrm{\Gamma}_{\mathrm{H}}P_{\mathrm{H}}+\xi_{\mathrm{H}}
\label{eq:PHeq}\\
\dot{P}_{\mathrm{K}}&=\gamma_{\mathrm{K}}B\times P_{\mathrm{K}}+\mathrm{\Gamma}_{\mathrm{KH}}P_{\mathrm{H}}-\mathrm{\Gamma}_{\mathrm{K}}P_{\mathrm{K}}+\mathrm{\Gamma}_{\mathrm{P}}\hat{z}+\xi_{\mathrm{K}}
\label{eq:PKeq}
\end{align}

In the proposed configuration, the magnetic field is given by  $B=B_\perp \mathrm{cos}(\omega t)\hat{x}+B_z\hat{z}$. We define the spin polarization in the rotating frame $P_\mathrm{i}^+=P_\mathrm{i}^x+iP_\mathrm{i}^y$ and the noise term $\xi_\mathrm{i}^+=\xi_\mathrm{i}^x+i\xi_\mathrm{i}^y$  and can cast the four equations of motion of the different components by ~(\ref{eq:PHplus}-~\ref{eq:PKz}).

We transform the spin-polarization to a coordinates system that rotates at the driving field  $\omega$, as given by 

\begin{align}
P'^+_{\mathrm{i}}&=P^+_{\mathrm{i}} e^{-i\omega t};~P'^z_{\mathrm{i}}=P^z_{\mathrm{i}};~\xi'^+_{\mathrm{i}}=\xi^+_{\mathrm{i}} e^{-i\omega t}\\
\dot{P}'^+_{\mathrm{i}}&=\dot{P}^+_{\mathrm{i}}e^{-i\omega t}-i\omega P^+_{\mathrm{i}} e^{-i\omega t}
\label{eq:Ptrans}
\end{align}

Where the new noise variables are statistically equivalent to the original ones. We note that the prime indicating the rotating frame was omitted from the main text for brevity. We now neglect fast rotating terms, equivalent to keeping the low-frequency response measured in a Lock-In amplifier, yielding ~(\ref{eq:PHplusdot}-~\ref{eq:PKzdot}). Then we calculate the steady state of the polarization in the rotating frame, which is given as the solution to ~(\ref{eq:Matrxform}).

For a weak Larmor driving amplitude $\gamma_{\mathrm{K}}B_\perp<<\Gamma_{\mathrm{K}}$ and  $\gamma_{\mathrm{K}}B_\perp  \lesssim\Gamma_{\mathrm{H}}$, we find that the axial polarizations are given by 
\begin{align}
P'^z_{\mathrm{H}}=&\frac{\Gamma_{\mathrm{H}}\Gamma_{\mathrm{p}}}{\Gamma_{\mathrm{HK}}\Gamma_{\mathrm{KH}}-\Gamma_{\mathrm{H}}\Gamma_{\mathrm{K}}} \\
P'^z_{\mathrm{K}}=&\frac{\Gamma_{\mathrm{HK}}\Gamma_{\mathrm{p}}}{\Gamma_{\mathrm{HK}}\Gamma_{\mathrm{KH}}-\Gamma_{\mathrm{H}}\Gamma_{\mathrm{K}}}
\label{eq:PZhk}
\end{align}For the case of potassium $(I=3/2)$ and hydrogen $(I=1/2)$, we use the relation: $\gamma_{\mathrm{H}}=2\gamma_{\mathrm{K}}$ and the fact that for the proposed working point the following conditions are satisfied:  $\Gamma_{\mathrm{H}}\ll\Gamma_{\mathrm{KH}}$ and  $(\gamma_{\mathrm{H}}-\omega)\ll\Gamma_{\mathrm{KH}}$. Then, the full solution for the transverse polarization terms is given by ~(\ref{eq:PHplussolution}) and ~(\ref{eq:PKplussolution}).

\begin{widetext}

\begin{align}
\dot{P}^+_{\mathrm{H}}&= -\mathrm{\Gamma}_{\mathrm{H}}P^+_{\mathrm{H}} +\mathrm{\Gamma}_{\mathrm{HK}}P^+_{\mathrm{K}}-i\gamma_{\mathrm{H}}B_{\perp}\mathrm{cos}(\omega t)P^z_{\mathrm{H}}+i\gamma_{\mathrm{H}}B_zP^+_{\mathrm{H}}+\xi_{\mathrm{H}}^+
\label{eq:PHplus}\\
\dot{P}^z_{\mathrm{H}}&= -\mathrm{\Gamma}_{\mathrm{H}}P^z_{\mathrm{H}} +\mathrm{\Gamma}_{\mathrm{HK}}P^z_{\mathrm{K}}-i\gamma_{\mathrm{H}}B_{\perp}\mathrm{cos}(\omega t)P^y_{\mathrm{H}}
\label{eq:PHz}\\
\dot{P}^+_{\mathrm{K}}&= -\mathrm{\Gamma}_{\mathrm{K}}P^+_{\mathrm{K}} +\mathrm{\Gamma}_{\mathrm{KH}}P^+_{\mathrm{H}}-i\gamma_{\mathrm{K}}B_{\perp}\mathrm{cos}(\omega t)P^z_{\mathrm{K}}+i\gamma_{\mathrm{K}}B_zP^+_{\mathrm{K}}+\xi_{\mathrm{K}}^+
\label{eq:PKplus}\\
\dot{P}^z_{\mathrm{K}}&= -\mathrm{\Gamma}_{\mathrm{K}}P^z_{\mathrm{K}} +\mathrm{\Gamma}_{\mathrm{KH}}P^z_{\mathrm{H}}-i\gamma_{\mathrm{K}}B_{\perp}\mathrm{cos}(\omega t)P^y_{\mathrm{K}}+\Gamma_{\mathrm{p}}
\label{eq:PKz}
\end{align}

\begin{align}
\dot{P}_{\mathrm{H}}'^+= & -\mathrm{\Gamma}_{\mathrm{H}}P'^+_{\mathrm{H}} +\mathrm{\Gamma}_{\mathrm{HK}}P'^+_{\mathrm{K}}-i\gamma_{\mathrm{H}}B_{\perp}P'^z_{\mathrm{H}}/2+i(\gamma_{\mathrm{H}}B_z-\omega)P'^+_{\mathrm{H}}+\xi_{\mathrm{H}}'^+
\label{eq:PHplusdot}\\  
\dot{P}'^z_{\mathrm{H}}= &-\mathrm{\Gamma}_{\mathrm{H}}P'^z_{\mathrm{H}} +\mathrm{\Gamma}_{\mathrm{HK}}P'^z_{\mathrm{K}}-i\gamma_{\mathrm{H}}B_{\perp}P'^+_{\mathrm{H}}/2
\label{eq:PHzdot}\\
\dot{P}'^+_{\mathrm{K}}= &-\mathrm{\Gamma}_{\mathrm{K}}P'^+_{\mathrm{K}} +\mathrm{\Gamma}_{\mathrm{KH}}P'^+_{\mathrm{H}}-i\gamma_{\mathrm{K}}B_{\perp}P'^z_{\mathrm{K}}/2+i(\gamma_{\mathrm{K}}B_z-\omega)P'^+_{\mathrm{K}}+\xi_{\mathrm{K}}^+
\label{eq:PKplusdot}\\
\dot{P}'^z_{\mathrm{K}}= &-\mathrm{\Gamma}_{\mathrm{K}}P'^z_{\mathrm{K}} +\mathrm{\Gamma}_{\mathrm{KH}}P'^z_{\mathrm{H}}-i\gamma_{\mathrm{K}}B_{\perp}P'^+_{\mathrm{K}}/2+\Gamma_{\mathrm{p}}
\label{eq:PKzdot}
\end{align}

\begin{align} \label{eq:Matrxform}
\begin{pmatrix}
-\Gamma_{\mathrm{H}}+i(\gamma_{\mathrm{H}}B_z-\omega)& -i\gamma_{\mathrm{H}}B_\perp/2 & \Gamma_{\mathrm{HK}}&0\\
-i\gamma_{\mathrm{H}}B_\perp/2&-\Gamma_{\mathrm{H}} & 0 &\Gamma_{\mathrm{HK}}\\
\Gamma_{\mathrm{KH}}& 0 & -\Gamma_{\mathrm{K}}+i(\gamma_{\mathrm{K}}B_z-\omega)&-i\gamma_{\mathrm{K}}B_\perp/2\\
0& \Gamma_{\mathrm{KH}} & -i\gamma_{\mathrm{K}}B_\perp/2 &-\Gamma_{\mathrm{K}}\
\end{pmatrix}
\begin{pmatrix}
P'^+_{\mathrm{H}}\\
P'^z_{\mathrm{H}}\\
P'^+_{\mathrm{K}}\\
P'^z_{\mathrm{K}}\
\end{pmatrix}=
\begin{pmatrix}
\xi'^+_{\mathrm{H}}\\
0\\
\xi'^+_{\mathrm{K}}\\
\Gamma_{\mathrm{p}}\
\end{pmatrix}
\end{align}

\begin{align}
P'^+_{\mathrm{H}}=
\frac{\Gamma_{\mathrm{K}}-i(\gamma_{\mathrm{K}}B_z-\omega)}
{\Gamma_{\mathrm{HK}}\Gamma_{\mathrm{KH}}-
(\Gamma_{\mathrm{K}}-i(\gamma_{\mathrm{K}}B_z-\omega))
(\Gamma_{\mathrm{H}}-i(\gamma_{\mathrm{H}}B_z-\omega))}\left(i\gamma_{\mathrm{K}}B_\perp\Gamma_{\mathrm{p}}P'^z_{\mathrm{H}}+
\xi'^+_{\mathrm{H}}+\frac{\Gamma_{\mathrm{HK}}\xi'^+_{\mathrm{K}}}{\Gamma_{\mathrm{K}}-i(\gamma_{\mathrm{K}}B_z-\omega)}\right)
\label{eq:PHplussolution}
\end{align}
\begin{align}
P'^+_{\mathrm{K}}=
\frac{\Gamma_{\mathrm{H}}-i(\gamma_{\mathrm{H}}B_z-\omega)}
{\Gamma_{\mathrm{HK}}\Gamma_{\mathrm{KH}}-
(\Gamma_{\mathrm{K}}-i(\gamma_{\mathrm{K}}B_z-\omega))
(\Gamma_{\mathrm{H}}-i(\gamma_{\mathrm{H}}B_z-\omega))}\left(\frac{i\gamma_{\mathrm{K}}B_\perp\Gamma_{\mathrm{HK}}\Gamma_{\mathrm{KH}}P'^z_{\mathrm{K}}}{\Gamma_{\mathrm{H}}(\Gamma_{\mathrm{H}}-i(\gamma_{\mathrm{H}}B_z-\omega))}+ \xi'^+_{\mathrm{K}}+
\frac{\Gamma_{\mathrm{KH}}\xi'^+_{\mathrm{H}}}{\Gamma_{\mathrm{H}}-i(\gamma_{\mathrm{H}}B_z-\omega)}\right)
\label{eq:PKplussolution}
\end{align} 
\end{widetext}

\subsection{A typical spin projection noise for the potassium}
We can use the previous equations and construct the average response, whose amplitude is given by ~(\ref{eq:NormSigSQ}) with a peak response at the resonance ~(\ref{eq:NormSigSQres}). We next calculate the noise spectrum associated with atom-projection-noise. We consider fluctuations from the average value ~(\ref{eq:PKplussolNoise}) and  consider the time averaged variance at low frequencies ~(\ref{eq:PnoiseTime}). In the main text, we considered the scaling of the response at the resonance frequency of the hydrogen $\gamma_{\mathrm{H}}B_z=\omega$ under the assumption that  $\Gamma_{\mathrm{K}}\sim\Gamma_{\mathrm{HK}}\sim\gamma_{\mathrm{K}}B_\perp$.
To derive the magnetic noise level associated with atom projection noise, we linearize the response function in ~(\ref{eq:NormSigSQ}) as a function of $\gamma_{\mathrm{H}}B_z-\omega$ and find compare with the variance in ~(\ref{eq:PnoiseTime}) , yielding  Eq.~(\ref{eq:FMN}) in the main text. 

To construct Fig.~(\ref{fig:optimal}) in the main text, for each combination of number densities $n_{\mathrm{H}}$, $n_{\mathrm{K}}$ we search for a solution that minimizes  at the hydrogen resonance, using the optical pumping rate $\Gamma_{\mathrm{p}}$ and the amplitude of the magnetic drive $B_{\perp}$ as the search parameters.  Then, the magnetic noise level is given simply by
\begin{align}
\delta B_{\mathrm{APN}}=\left|\delta B_z\right|\sqrt{\frac{\Delta P'^2_{\mathrm{N}}}{\left<P'^+_{\mathrm{K}}\right>\left<P'^-_{\mathrm{K}}\right>}}
\label{eq:deltaBAPN}
\end{align}
Using the noise statistics
\begin{align}
\left<\xi_{qi}\left(t\right)\xi_{pj}\left(t'\right)\right>_{\pm}=\delta_{qp}\delta_{ij}\delta\left(t-t'\right)\frac{\sqrt{2}\Gamma_{q}}{n_q V} \label{eq:NoiseStat}
\end{align}
we compute the noise spectrum and find ~\ref{eq:PnoiseTimeQ}.

\begin{widetext}
\begin{align}
\left<P'^+_{\mathrm{K}}\right>\left<P'^-_{\mathrm{K}}\right>=
\frac{\left(P'^z_{\mathrm{K}}\gamma_{\mathrm{K}}B_\perp\right)^2}
{\Gamma^2_{\mathrm{H}}\left(1+
2\frac{\left(\gamma_{\mathrm{K}}B_z-\omega\right)\left(\gamma_{\mathrm{H}}B_z-\omega\right)-\Gamma_{\mathrm{H}}\Gamma_{\mathrm{K}}}
{\Gamma_{\mathrm{HK}}\Gamma_{\mathrm{KH}}}+
\frac{\Gamma_{\mathrm{K}}^2
+\left(\gamma_{\mathrm{K}}B_z-\omega\right)^2}
{\Gamma_{\mathrm{KH}}^2}
\frac{\Gamma_{\mathrm{H}}^2
+\left(\gamma_{\mathrm{H}}B_z-\omega\right)^2}
{\Gamma_{\mathrm{HK}}^2}
\right)}
\label{eq:NormSigSQ}
\end{align}

\begin{align}
\left<P'^+_{\mathrm{K}}\left(\gamma_{\mathrm{H}}B_z\right)\right>\left<P'^-_{\mathrm{K}}\left(\gamma_{\mathrm{H}}B_z\right)\right>=
\left(\frac{P'^z_{\mathrm{K}}\gamma_{\mathrm{K}}B_\perp}{\Gamma_{\mathrm{H}}}\right)^2
\frac{1}{\left(1-\frac{\Gamma_{\mathrm{H}}\Gamma_{\mathrm{K}}}{\Gamma_{\mathrm{HK}}\Gamma_{\mathrm{KH}}}\right)^2+
\left(\frac{\gamma_{\mathrm{K}}B_z\Gamma_{\mathrm{H}}}{\Gamma_{\mathrm{HK}}\Gamma_{\mathrm{KH}}}\right)^2}
\label{eq:NormSigSQres}
\end{align}

\begin{align}
\Delta P'^+_{\mathrm{K}}=
\frac{\Gamma_{\mathrm{H}}-i(\gamma_{\mathrm{H}}B_z-\omega)}
{\Gamma_{\mathrm{HK}}\Gamma_{\mathrm{KH}}-
(\Gamma_{\mathrm{K}}-i(\gamma_{\mathrm{K}}B_z-\omega))
(\Gamma_{\mathrm{H}}-i(\gamma_{\mathrm{H}}B_z-\omega))}\left(\xi'^+_{\mathrm{K}}+
\frac{\Gamma_{\mathrm{KH}}\xi'^+_{\mathrm{H}}}{\Gamma_{\mathrm{H}}-i(\gamma_{\mathrm{H}}B_z-\omega)}\right)
\label{eq:PKplussolNoise}
\end{align} 

\begin{align}
\Delta P'^2_{\mathrm{N}}\left(\omega\right)=
\left<\Delta P'^+_{\mathrm{K}}\left(t\right)\Delta P'^-_{\mathrm{K}}\left(t\right)\right>=
\left<\int\int \Delta P'^+_{\mathrm{K}}\left(t\right)\Delta P'^-_{\mathrm{K}}\left(t'\right)dt'dt\right>
\label{eq:PnoiseTime}
\end{align} 

\begin{align}
\frac{\Delta P'^2_{\mathrm{N}}\left(\omega\right)}{P'^+_{\mathrm{K}}\left(\omega\right)P'^-_{\mathrm{K}}\left(\omega\right)}=
\frac{2t\Gamma^2_{\mathrm{H}}\left(\Gamma^2_{\mathrm{H}}+\left(\gamma_{\mathrm{H}}B_z-\omega\right)^2\right)}
{V\left(P'^z_{\mathrm{K}}\gamma_{\mathrm{K}}B_\perp\right)^2\Gamma_{\mathrm{HK}}^2\Gamma_{\mathrm{KH}}^2}
\left(\frac{\Gamma_{\mathrm{K}}}{n_{\mathrm{K}}}+\frac{\Gamma_{\mathrm{KH}}^2}
{\Gamma_{\mathrm{H}}^2+\left(\gamma_{\mathrm{H}}B_z-\omega\right)^2}
\frac{\Gamma_{\mathrm{H}}}
{n_{\mathrm{H}}}\right)
\label{eq:PnoiseTimeQ}
\end{align}
\end{widetext}

\section{Decoherence by magnetic field gradient} \label{dB_restrictions}

In this appendix, we numerically analyze the decoherence induced by magnetic field inhomogeneity in the proposed experimental configuration, aiming to characterize the decoherence of hydrogen atoms due to magnetic field gradient. We consider a spherical cell with a diameter of $D=25$ mm, operating at a temperature of $80^{\circ}\textrm{C}$. 
The magnetic field within the cell is assumed to follow the form $B_z=B_0+(2z/D)\delta B$, where $B_0$ is constant at $0.4,\textrm{G}$, and a co-linear magnetic gradient of amplitude $\delta B$ is present. 

To characterize the evolution, we assume that hydrogen atoms move ballistically inside the spherical enclosure, undergoing simple bouncing off the walls. Fig.~\ref{fig:Bgrad_simulation}a illustrates the trajectory of an atom over time for the first $50$ collisions. In Fig.~\ref{fig:Bgrad_simulation}b, we demonstrate the local field probed by the atoms during their flight, considering two different trajectories and magnetic field values. Despite short-term field variations, atoms average over the magnetic field within the cell over longer durations. To quantify this averaging, we conducted $N=100$ numerical simulations for each $\delta B$, each extending beyond $7\times10^4$ wall collisions. We employed fine time steps and ensured convergence of the results in the simulation parameters.\par
The magnetic precession of the $i$th atom along the $x$ direction is represented by ~(\ref{eq:MagPrecess})
\begin{align} \label{eq:MagPrecess}
S_{x}^{(i)}(t)=\sin\left(\gamma B_0 t + \gamma\int_{0}^{t}  \delta B_i(t')dt'\right)
\end{align}
Here $\delta B_i(t)\equiv (2z_i(t)/D)\delta B$, with $z_i(t)$ denoting the position of the $i$th atom. We implicitly assume that all spins start precession in phase (e.g., following optical pumping in the $y$ direction). The ensemble-averaged signal, $S_\textrm{H}(t)\equiv\sum_{i=1}^NS_{x}^{(i)}(t)$, is then fitted to the function $\sin(\gamma B_0 t)e^{-t/T_\textrm{H}^{(\nabla B)}}$, with $\gamma$ representing the hydrogen gyromagnetic ratio and $T_\textrm{H}^{(\nabla B)}$ treated as a free fitting parameter. Fig.~\ref{fig:Bgrad_simulation}c displays $S_\textrm{H}$ for two configurations, with the corresponding fit values of $T_\textrm{H}^{(\nabla B)}$ shown in Fig.~\ref{fig:Bgrad_simulation}. In the considered regime, we observe that the decoherence rate induced by the magnetic field gradient scales as 
$\Gamma_H^{(\nabla B)}\equiv 1/T_{\textrm{H}}^{(\nabla B)}\cong \eta(\nabla B)^2$ with $\eta=85 ~\mathrm{cm}^2\mathrm{mG}^{-2}\mathrm{s}^{-1}$.

\bibliography{refs}

\end{document}